\documentstyle[12pt]{article}
\def\lsim{\mathrel{\rlap{\raise 2.5pt \hbox{$<$}}\lower 2.5pt}}
\def\gsim{\mathrel{\rlap{\raise 2.5pt \hbox{$>$}}\lower 2.5pt}}
\begin{document}
\bibliographystyle{plain}
\thispagestyle{empty}
\vspace{-5mm}
\begin{small}
\begin{flushright}
DESY 99-019\\ % [-2.5mm]
IISc-CTS-3/99\\ % [-2.5mm]
hep-ph/9902405\\ % [-2.5mm]
February 1999
\end{flushright}
\end{small}
\begin{center}
{\bf  Infra-red stable fixed points of R-parity violating Yukawa couplings in
supersymmetric models}
\vskip 2.5cm
B. Ananthanarayan, \\
\begin{small}
Centre for Theoretical Studies, 
Indian Institute of Science,\\ 
Bangalore 560 012, India\\ 
\end{small}
\medskip
P. N. Pandita,\\
\begin{small}
Theory Group, Deutsches Elektronen-Synchrotron DESY,\\
Notkestrasse 85, D 22603 Hamburg, Germany\\
and\\
Department of Physics, North-Eastern Hill University,\\
Shillong 793 022, India\footnote{Permanent Address}
\end{small}
\end{center}
\vskip 2.25cm
\begin{abstract}
We investigate the infra-red stable fixed points of the Yukawa couplings
in the minimal version of the supersymmetric standard model
with R-parity violation.
Retaining only the  R-parity violating couplings of  higher
generations, we analytically study the solutions of the renormalization group
equations of these couplings
together with the top- and b-quark Yukawa couplings.
We show that
only the B-violating coupling $\lambda^{''}_{233}$
approaches a non-trivial
infra-red stable fixed point, whereas all other non-trivial
fixed point solutions are  either unphysical or unstable in the
infra-red region. However, this fixed point solution predicts a top-quark
Yukawa coupling which is incompatible with the 
top quark mass for any value of $\tan\beta$.
\end{abstract}
\newpage

There is considerable interest in the study of infra-red~(IR)
stable fixed points of the standard model~(SM) and its extensions, especially
those of the minimal supersymmetric standard model~(MSSM).  This interest
follows from the fact that in the SM~(and in the MSSM) there are large
number of unknown dimensionless Yukawa couplings, as a consequence of
which the fermion masses cannot be predicted.  One may attempt to relate
the Yukawa couplings to the gauge couplings via the Pendleton-Ross
infra-red stable fixed point~(IRSFP) for the top-quark Yukawa
coupling~\cite{pendross}, or via the quasi-fixed point 
behaviour~\cite{hill}.    
The predictive power of the SM and its supersymmetric extensions may, thus,
be enhanced if the renormalization group~(RG) running of the parameters
is dominated by IRSFPs. Typically, these fixed points are for
ratios like Yukawa coupling to the gauge coupling, or, in the context of 
supersymmetric models, the supersymmetry breaking tri-linear A-parameter
to the gaugino mass, etc.  These ratios do not always attain their fixed
points values at the weak scale, the range between the GUT~(or Planck)
scale and the weak scale being too small for the ratios to closely
approach the fixed point.  Nevertheless, the couplings may
be determined by quasi-fixed point behaviour~\cite{hill}, where the
value of the Yukawa coupling at the weak sale is independent of
its value at the GUT scale, provided the Yukawa couplings at the unification
scale are large.  For the fixed point or quasi-fixed point scenarios
to be successful, it is necessary that
these fixed points be stable~\cite{allanach, abel, jack}.  

Since supersymmetry~\cite{nilles} necessciates the introduction of
superpartners for all known particles in the SM (in addition to the
introduction of two Higgs doublets), which transform in an indentical
manner under the gauge group, we have additional Yukawa couplings in 
supersymmetric models which violate~\cite{weinberg} baryon number~(B)
or lepton number~(L).  In the MSSM a discrete symmetry called
R-parity~($R_p$) is invoked to eliminate these B and L violating 
Yukawa couplings~\cite{farrar}.  However, the assumption of $R_p$
conservation at the level of MSSM appears to be {\it ad hoc}, since it is
not required for the internal consistency of the model. Therefore, the 
study of MSSM, including R-parity violation, deserves a serious 
consideration.

Recently attention has been focussed on the study of renormalization
group evolution of $R_p$ violating Yukawa couplings
of the MSSM~\cite{barger}, 
and their quasi-fixed points.
This has led to certain insights and constraints on the quasi-fixed
point behavior of some of the $R_p$ violating Yukawa couplings, involving
higher generation indices.   We  recall that the usefulness of
the fixed point and quasi-fixed point scenarios is the existence 
of {\it stable} infra-red fixed points.
The purpose of this paper is to address the important
question of the infra-red fixed points of supersymmetric models 
with $R_p$ violation, and their stability. Our  interest is in the 
structure of the infra-red stable fixed points, rather than the actual
values of the fixed points.

To this end we shall consider the supersymmetric
standard model with the minimal particle content and with $R_P$ violation,
and refer to it as MSSM with R-parity violation.
We begin by recalling some of the basic features of the model.
The superpotential of the MSSM is given by 
\begin{equation}\label{mssmsuperpotential}
W=\mu H_1 H_2 + (h_u)_{ab} Q^a_L \overline{U}^b_R
 H_2 + (h_d)_{ab} Q^a_L \overline{D}^b_R H_1 + (h_E)_{ab} L^a_L 
 \overline{E}^b_R H_1,
\end{equation}
to which we add the L and B violating terms
\begin{eqnarray}
& \displaystyle W_L=\mu_i L_i H_2  + {1\over 2}\lambda_{abc} L^a_L L^b_L 
\overline{E}^c_R
+ \lambda'_{abc} L^a_L Q^b_L\overline{D}^c_R, & \label{Lviolating}\\
& \displaystyle W_B={1\over 2}\lambda''_{abc} \overline{D}^a_R
\overline{D}^b_R \overline{U}^c_R,  & \label{Bviolating} 
\end{eqnarray}
respectively,  as allowed by gauge invariance and supersymmetry. 
In Eq.~(\ref{mssmsuperpotential}), $(h_U)_{ab}$, $(h_D)_{ab}$
and $(h_E)_{ab}$ are the Yukawa coupling matrices, with $a,\, b,\, c$
as the generation indices.  
The Yukawa couplings
$\lambda_{abc}$ and $\lambda''_{abc}$ are antisymmetric in their
first two indices due to $SU(2)_L$ and $SU(3)_C$ group
structure.  Phenomenological
studies of supersymmetric models of this type
have placed  constraints~\cite{dreiner} on the various couplings
$\lambda_{abc}$, $\lambda'_{abc}$ and $\lambda''_{abc}$, but there
is still considerable room left.  We note that the simultaneous presence
of the terms in Eq.~(\ref{Lviolating}) and Eq.~(\ref{Bviolating})
is essentially ruled out by the stringent constraints~\cite{smirnov}
implied by the lack of
observation of nucleon decay.

In addition to the dominant third generation Yukawa couplings
$h_t \equiv (h_U)_{33}$, $h_b \equiv (h_D)_{33}$ and $h_{\tau} \equiv
(h_E)_{33}$
in the superpotential~(\ref{mssmsuperpotential}), there are 36 independent
$R_p$ violating couplings $\lambda_{abc}$ and  $\lambda'_{abc}$
in Eq.~(\ref{Lviolating}),  and 9 independent $\lambda''_{abc}$
in Eq.~(\ref{Bviolating}).  Thus, one would have to solve 39
coupled non-linear evolution equations in the L-violating case,
and 12 in the B-violating case, in order to study 
the evolution of the Yukawa couplings in the minimal model with $R_p$ violation.
In order to render the Yukawa coupling evolution tractable, we
need to make certain plausible simplications.  Motivated by
the generational hierarchy of the conventional Higgs couplings, 
we shall assume that an analogous hierarchy amongst the different
generations of $R_p$ violating couplings exists.  Thus we shall
retain only the couplings $\lambda_{233}$, $\lambda'_{333}$ and
$\lambda''_{233}$, 
and neglect the rest.  We note that
the $R_p$ violating couplings to higher generations evolve more
strongly because of larger Higgs couplings in their evolution equations,
and hence could take larger values than the corresponding couplings to
the lighter generations.  Furthermore, the experimental upper limits are
stronger for the $R_p$ violating Yukawa couplings corresponding
to the lighter generations.

We shall first consider the evolution of Yukawa couplings
arising from superpotentials~(\ref{mssmsuperpotential}) and (\ref{Bviolating}),
which involve baryon number violation.
The one-loop renormalization group equations for $h_t,\, h_b, \,
h_\tau$ and $\lambda''_{233}$ (all others set to zero) are:
\begin{eqnarray} 
& \displaystyle 16 \pi^2 {dh_t\over d(\ln\, \mu)}=
h_t\left(6 h_t^2 + h_b^2+2\lambda''^2_{233}-{16\over 3} g_3^2
-3 g_2^2 -{13\over 15} g_1^2\right), & \nonumber \\
& \displaystyle 16 \pi^2 {dh_b\over d(\ln\, \mu)}=
h_b\left(  h_t^2 +6h_b^2+h_\tau^2+2\lambda''^2_{233}-{16\over 3} g_3^2
-3 g_2^2 -{7\over 15} g_1^2\right), & \nonumber \\
& \displaystyle 16 \pi^2 {dh_\tau\over d(\ln\, \mu)}=
h_\tau\left(  3 h_b^2+4 h_\tau^2
-3 g_2^2 -{9\over 5} g_1^2\right), & \label{htauequation} \\
& \displaystyle 16 \pi^2 {d\lambda''_{233}\over d(\ln\, \mu)}=
\lambda''_{233}\left(2 h_t^2 +2h_b^2+6\lambda''^2_{233}-8 g_3^2
-{4\over 5} g_1^2\right). & \nonumber 
\end{eqnarray}
For completeness we list the well-known evolution equations
for the gauge couplings, which at one-loop order are identical to
those in the MSSM, since the additional Yukawa coupling(s) do not play
a role at this order: 
\begin{eqnarray}
& \displaystyle 16 \pi^2{d g_i\over d(\ln\mu)}=b_i g_i^3, \, i=1,2,3, &
\end{eqnarray}
with $b_1=33/5,\, b_2=1, \, b_3=-3$. 
With the definitions
\begin{equation}\label{redefinitions}
R_t={h_t^2\over g_3^2}, \,\,\, R_b={h_b^2\over g_3^2}, \,\,\,
R_\tau={h_\tau^2\over g_3^2},  \,\,\, R''={\lambda''^2_{233}\over g_3^2},
\end{equation}
and retaining only the $SU(3)_C$ gauge coupling constant, we can
rewrite the renormalization group equations as ($\tilde{\alpha}_3
=g_3^2/(16\pi^2)$):
\begin{eqnarray}
& \displaystyle {dR_t\over dt}=\tilde{\alpha_3} R_t
\left[\left({16\over 3}+b_3\right)-
6 R_t - R_b - 2 R''\right],  & \label{rtequation}
 \\
& \displaystyle {dR_b\over dt}=\tilde{\alpha_3} R_b
\left[\left({16\over 3}+b_3\right)
- R_t -6 R_b -R_\tau- 2 R''\right],  & \label{rbequation}
 \\
& \displaystyle {dR_\tau\over dt}=\tilde{\alpha_3} R_\tau
\left[b_3 -3 R_b -4 R_\tau\right],  & \label{rtauequation}
\\
& \displaystyle {dR''\over dt}=\tilde{\alpha_3} R''
\left[\left(8+b_3\right)
-2 R_t -2 R_b - 6 R''\right], & \label{rdoubleprimeequation}
\end{eqnarray}
where $b_3=-3$ is the beta function for $g_3$ in the MSSM, and
$t=-\ln\, \mu^2$.  
Ordering the ratios as  $R_i = (R'', R_{\tau}, R_b, R_t)$, 
we rewrite the RG equations 
(\ref{rtequation}) - (\ref{rdoubleprimeequation})
in the form~\cite{allanach}
\begin{equation}\label{componentequation}
{d R_i\over dt}=\tilde{\alpha}_3 R_i
\left[(r_i+b_3)-\sum_j S_{ij} R_j\right],
\end{equation}
where $r_i=\sum_R 2 \, C_R$,  $C_R$ is the QCD Casimir
for the various fields ($C_Q=C_{\overline{U}}=C_{\overline{D}}=4/3$), 
the sum is over the representation of
the three fields associated with the trilinear coupling that enters
$R_i$, and $S$ is a matrix whose value is fully specified by
the wavefunction anomalous dimensions. 
A fixed point is then reached
when the right hand side of Eq.~(\ref{componentequation}) is
0 for all $i$.  If we were to write the fixed-point solutions as
$R_i^*$, then there are two fixed point values for each coupling:
$R_i^*=0$, or
\begin{equation}
\left[\left(r_i+b_3\right) -\sum_j S_{ij} R_j^* \right]=0.
\end{equation}
It follows that the non-trivial fixed point solution is
\begin{equation}\label{ristarequation}
R_i^*=\sum_j (S^{-1})_{ij} (r_j+b_3).
\end{equation}
Since we shall consider the fixed points of the couplings $h_t$, $h_b$ and
$\lambda_{233}^{''}$ only, we shall ignore the evolution equation
(\ref{rtauequation}). 
However,  the coupling $h_\tau$ does enter
the evolution Eq.~(\ref{rbequation}) of
$h_b$, but it can be related to $h_b$ at the weak scale (which we 
take to be the top-quark mass),  since
\begin{equation}\label{htaurelation}
h_\tau(m_t)={\sqrt{2} m_\tau(m_t)\over \eta_\tau v \cos \beta},
\end{equation}
and
\begin{equation}\label{2ndhtaurelation}
h_\tau(m_t)={m_\tau(m_\tau)\over m_b(m_b)} {\eta_b \over \eta_\tau} h_b(m_t)
= 0.6 h_b(m_t),
\end{equation}
where $\eta_b$ gives the QCD or QED running~\cite{barger2} 
of the b-quark mass $m_b(\mu)$ between $\mu = m_b$ and $\mu = m_t$
(similarly for $\eta_{\tau}$), 
and $\tan\beta=v_2/v_1$ is the usual ratio of the Higgs vacuum expectation
values in the MSSM, with   $v=(\sqrt{2} G_F)^{-1/2}=246$ GeV.
The anomalous dimension matrix $S$ can, then, be written as 
\begin{equation}\label{Smatrix}
S=\left(
\begin{array}{ c c c }
6 & 2 & 2 \\
2 & 6 + \eta & 1 \\
2 & 1 & 6 \\
\end{array}
\right),
\end{equation}
where $\eta=h_\tau^2(m_t)/h_b^2(m_t)\simeq 0.36$ is the factor
coming from Eq.~(\ref{2ndhtaurelation}). We, therefore, get
the following fixed point solution for the ratios: 
\begin{eqnarray}
& \displaystyle R_1^* \equiv R''^*={385+76 \eta \over 3(170 + 32 \eta)}
\simeq 0.76, & \nonumber \\
& \displaystyle R_2^* \equiv R_b^*={20\over 170+ 32 \eta}
\simeq 0.11, & \label{rstarsolution} \\
& \displaystyle R_3^* \equiv R_t^*={20+4 \eta \over 170 + 32 \eta}\simeq 0.12. & \nonumber
\end{eqnarray}
Since each of the $R_i$'s is positive, this is a theoretically acceptable
fixed point solution.

We next try to find a fixed point solution with $R''^*=0$, with  
$R_b$ and $R_t$ being given by their non-zero solutions. We 
need to consider only the lower right hand $2\times 2$ sub-matrix
of the matrix S in Eq.~(\ref{Smatrix}) to obtain the fixed point
solutions for $R_b$ and $R_t$ in this case.  We then have
\begin{eqnarray}
& \displaystyle R_1^* \equiv R''^* = 0, & \nonumber \\
& \displaystyle R_2^*\equiv R_b^* ={35\over 3 \, (35+6\eta)} \simeq 0.36,
 &  \label{2ndrstarsolution} \\
& \displaystyle R_3^* \equiv R_t^*={7(5+ \eta) \over 3 (35+6\eta)} \simeq
0.34. & \nonumber
\end{eqnarray}
This is also a theoretically acceptable solution, as all the fixed point
values are non-negative.  We must also consider the fixed point with
$R_b^*=0$, which is relevant for the low values of the parameter
$\tan\beta.$  In this case, we have to reorder  the couplings as 
$R_i = (R_b,\, R'', R_t)$, so that we have the anomalous dimension
matrix (in this case denoted as $\tilde{S}$)
\begin{equation}\label{newSmatrix}
\tilde{S}=
\left(
\begin{array}{c c c}
6+\eta & 2 & 1 \\
2 & 6 & 2 \\
1 & 2 & 6 \\
\end{array}
\right).
\end{equation}
Since $R_b^*=0,$ we have to determine the non-zero fixed point values for
$R''$ and $R_t$ only.  For this we consider the lower right hand 
$2\times 2$ submatrix of the matrix in (\ref{newSmatrix}) to obtain
\begin{eqnarray}
& \displaystyle R_1^*\equiv R_b^* =0, & \nonumber \\
& R_2^* \equiv R''^*={19\over 24} \simeq 0.79, & \label{3rdrstarsolution}\\
& \displaystyle R_3^*\equiv R_t^*={1\over 8}\simeq 0.12. & \nonumber 
\end{eqnarray}
which is an acceptable fixed point solution as well.
Since there are more than one theoretically acceptable IRSFPs in
this case, it is important to determine which, if any, is more
likely to be realized in nature.  To this end, we must examine the
stability of each of the fixed point solutions.

The infra-red stability of a fixed point solution is determined 
by the sign of the eigenvalues of the matrix $A$ whose
entries are ($i$ not summed over)~\cite{allanach}
\begin{equation}\label{aijmatrix}
A_{ij}={1\over b_3} R_i^* S_{ij},
\end{equation}
where $R_i^*$ is the set of the fixed point solutions of the
Yukawa couplings under consideration, and $S_{ij}$ is the matrix
appearing in the corresponding RG equations (\ref{componentequation})
for the ratios $R_i$.  For stability, we require all the eigenvalues
of the matrix Eq.~(\ref{aijmatrix}) to have negative real parts
(note that the QCD $\beta$-function $b_3$ is negative).  Considering
the fixed point solution (\ref{rstarsolution}), the matrix
$A$ can be written as
\begin{equation}\label{amatrix}
A=
-{1\over 3}\left(
\begin{array}{c c c}
6 R_1^* & 2 R_1^* & 2 R_1^* \\
2 R_2^* & (6+\eta) R_2^* & R_2^* \\
2 R_3^* & R_3^* & 6 R_3^* \\
\end{array}
\right),
\end{equation}
where $R_i^*$ are given in Eq.~(\ref{rstarsolution}).
The eigenvalues of the matrix Eq.~(\ref{amatrix}) are calculated
to be 
\begin{equation}\label{lambda1}
\lambda_1=-1.6, \, \lambda_2=-0.2, \, \lambda_3=-0.2,
\end{equation}
which shows that the fixed point (\ref{rstarsolution}) 
is an infra-red stable fixed point. We note that the eigenvalue
$\lambda_1$ is larger in magnitude as compared 
to the other eigenvalues   in (\ref{lambda1}), 
indicating that the fixed point for $\lambda_{233}^{''}$
is more attractive, and hence more relevant.

Next, we consider the stability of the fixed point solution 
(\ref{2ndrstarsolution}).  Since in this case the fixed point
of the coupling $R''^*=0$, we have to obtain the behaviour of this
coupling around the origin.  This behaviour is determined by the
eigenvalue~\cite{allanach}
\begin{equation}\label{secondevequation}
\lambda_1={1\over b_3}\left[ \sum_{j=2}^3 S_{1j} R_j^* -(r_1+b_3)\right],
\end{equation}
where $r_1=2(C_{\overline{U}}+C_{\overline{D}})=8$,  the
$C$s are the quadratic Casimirs of the fields occuring in the
B-violating terms in the superpotential (\ref{Bviolating}),
and the $S_{ij}$ is the matrix (\ref{Smatrix}), with the fixed points
$R_i^*,\, i=1,2,3$ given by Eq.~(\ref{2ndrstarsolution}).  
Inserting these values in Eq.~(\ref{secondevequation}), we find
\begin{equation}\label{lambda1equation}
\lambda_1={385+76 \eta \over 9 \, (35+6 \eta)} >0,
\end{equation}
thereby indicating that the fixed point is unstable in the infra-red.
The behaviour of the couplings $R_b$ and $R_t$ around their respective
fixed points is governed by the eigenvalues of the  
the $2\times 2$ lower submatrix of the matrix $A$ in Eq.~(\ref{amatrix})
\begin{equation}\label{anotheraijequation}
-{1\over 3}\left(
\begin{array}{c c}
(6+\eta) R_2^* & R_2^* \\
R_3^* & 6 R_3^* \\
\end{array}
\right),
\end{equation}
which we find to be 
\begin{equation} \label{anotherevequation}
\lambda_2=-0.78, \, \lambda_3=-0.56.
\end{equation}
Although $\lambda_2$ and $\lambda_3$ are negative, because of the result
(\ref{lambda1equation}), the fixed-point solution 
(\ref{2ndrstarsolution}) is unstable in the infra-red.
In other words, the $R_p$ conserving fixed point solution 
(\ref{2ndrstarsolution}) will never be achieved at low energies
and must be rejected.

Finally we come to the question of the stability of the fixed point
solution (\ref{3rdrstarsolution}).  The behaviour of the
coupling $R_b^*$ around the origin is determined by
the eigenvalue 
\begin{equation}\label{secondlambda1equation}
\lambda_1={1\over b_3}\left[\sum_{j=2}^3 \tilde{S}_{1j} R_j^* -(r_1+b_3)\right],
\end{equation}
where $r_1=2(C_{\overline{Q}}+C_{\overline{D}})=16/3$, and
$\tilde{S}$ is the matrix (\ref{newSmatrix}).  Inserting these
numbers, we find
\begin{equation}\label{thirdlambda1equation}
\lambda_1={5\over 24}\simeq 0.21 >0,
\end{equation}
with the other two eigenvalues for determining the stability given
by the eigenvalues of the matrix 
which  is obtained from the lower $2\times 2$ submatrix
of the matrix $\tilde{S}$ in (\ref{newSmatrix}). This submatrix
can be written as
\begin{equation}\label{secondaijmatrix}
\tilde{A} = -{1\over 3}\left(
\begin{array}{c c}
6 R_2^* & 2 R_2^* \\
2 R_3^* & 6 R_3^* \\
\end{array}
\right),
\end{equation}
where $R_2^*$ and $R_3^*$ are given by Eq.~(\ref{3rdrstarsolution}).
The eigenvalues are
\begin{equation}
\lambda_2=-1.61, \, \lambda_3=-0.22.
\end{equation}
It follows, once again, that the fixed point solution given in
(\ref{3rdrstarsolution}) is not stable in the infra-red and is,
therefore, never reached at low-energies.  

One may also consider  the case where the couplings
$\lambda''_{233}$ and $h_b$ attain trivial fixed point values, 
whereas $h_t$ attains a non-trivial fixed point value.
In this case we have $R^*_3\equiv R^*_t=7/18$, the well-known
Pendleton-Ross~\cite{pendross} top-quark
fixed point of the MSSM.
To  study the stability of this solution in the present context,
we must consider the eigenvalues
\begin{eqnarray}
& \displaystyle \lambda_i={1\over b_3}(S_{i3}R^*_3-(r_i+b_3)), \, 
i=1,2,  & \nonumber
\end{eqnarray}
where $S_{i3}$ are read off from the matrix (\ref{Smatrix}), which yields
\begin{eqnarray}
& \displaystyle \lambda_1={38\over 27}, \, \, \lambda_2={35\over 54}.
& \nonumber
\end{eqnarray}
Since the sign of each of $\lambda_1$ and $\lambda_2$ is positive,
this solution is also unstable in the infra-red region.
However, from our discussion of infra-red fixed
point solution (\ref{rstarsolution}),  it is clear
that the Pendelton-Ross fixed point would be
stable in case $h_b$ and $\lambda^{''}_{233}$ are small,  though
negligible at the GUT scale. In this case, these would, of course, 
evolve away from zero at the weak scale, though realistically they would 
still be small (but not zero) at the  weak scale.
Thus, the only true infra-red stable fixed
point solution is the baryon number, and  $R_p$,  violating solution 
(\ref{rstarsolution}).
This is one of the main conclusions of this paper.
We note that the value of $R_t^*$ in (\ref{rstarsolution}) is lower than
the corresponding value of $7/18$ in MSSM with $R_p$ conservation.

It is appropriate to examine the implications of
the value of $h_t(m_t)$ predicted by our
fixed point analysis for the top-quark mass. From (\ref{rstarsolution}),
and $\alpha_3(m_t)\simeq 0.1$,
the fixed point value for the top-Yukawa 
coupling is predicted to be $h_t(m_t) \simeq 0.4$. This translates into
a top-quark (pole) mass of about $m_t \simeq 70 \sin\beta$ GeV, 
which is incompatible with the measured value~\cite{topmass} of 
top mass, $m_t \simeq 174$ GeV, 
for any value of $\tan\beta$. 
It follows that the true fixed point obtained here
provides only a qualitative understanding of the top quark mass
in MSSM with $R_p$ violation.

We now turn to the study of the renormalization group evolution for
the lepton number violating,  and $R_p$,  violating couplings in the
superpotential (\ref{Lviolating}).  Here we shall consider
the dimensionless couplings $\lambda_{233}$ and $\lambda'_{333}$ only.
The relevant one-loop renormalization group equations are:
\begin{eqnarray} 
& \displaystyle 16 \pi^2 {dh_t\over d(\ln\, \mu)}=
h_t\left(6 h_t^2 + h_b^2+\lambda'^2_{333}-{16\over 3} g_3^2
\right), & \nonumber \\
& \displaystyle 16 \pi^2 {dh_b\over d(\ln\, \mu)}=
h_b\left(  h_t^2 +6h_b^2+h_\tau^2+6\lambda'^2_{333}-{16\over 3} g_3^2
\right), & \nonumber \\
& \displaystyle 16 \pi^2 {dh_\tau\over d(\ln\, \mu)}=
h_\tau\left(  3 h_b^2+4 h_\tau^2+4 \lambda_{233}^2+3\lambda'^2_{333}
\right), & \label{lephtauequation} \\
& \displaystyle 16 \pi^2 {d\lambda_{233}\over d(\ln\, \mu)}=
\lambda_{233}\left(4 h_\tau^2 +4 \lambda_{233}^2+3\lambda'^2_{333}
\right), & \nonumber \\
& \displaystyle 16 \pi^2 {d\lambda'_{333}\over d(\ln\, \mu)}=
\lambda'_{333}\left(h_t^2+6 h_b^2+ h_\tau^2 +\lambda_{233}^2+6\lambda'^2_{333}
-{16\over 3}g_3^2
\right). & \nonumber
\end{eqnarray}
Defining the new ratios
\begin{equation}\label{newratios}
R={\lambda_{233}^2\over g_3^2},  \,\,\,   R'={\lambda'_{333}\over g_3^2},
\end{equation}
we may now rewrite the equations (\ref{lephtauequation})  as 
\begin{eqnarray}
& \displaystyle {dR\over dt}=\tilde{\alpha_3} R
\left[b_3 -4 R -3 R' -4 R_\tau \right], & \label{requation} \\
& \displaystyle {dR'\over dt}=\tilde{\alpha_3} R'
\left[\left({16\over 3}+b_3\right)
- R -6 R' -R_\tau-6 R_b-R_t\right], & \label{rprimeequation} \\
& \displaystyle {dR_\tau\over dt}=\tilde{\alpha_3} R_\tau
\left[b_3 -4 R-3 R'-3 R_b -4 R_\tau\right],  & \label{leprtauequation}\\
& \displaystyle {dR_b\over dt}=\tilde{\alpha_3} R_b
\left[\left({16\over 3}+b_3\right)-6 R'
- R_\tau -6 R_b -R_t \right],  & \label{leprbequation}\\
& \displaystyle {dR_t\over dt}=\tilde{\alpha_3} R_t
\left[\left({16\over 3}+b_3\right)-R'
 -R_b -6 R_t \right].  & \label{leprtequation}
\end{eqnarray}
%\newpage
Ordering the ratios as $R_i = (R,R',R_\tau,R_b,R_t)$, we
can write these  RG equations  as:
\begin{equation}\label{lepcomponentequation}
{d R_i\over dt}=\tilde{\alpha}_3 R_i
\left[(r_i+b_3)-\sum_j S_{ij} R_j\right],
\end{equation}
where $r_i=\sum_R 2 C_R$, with $C_R$  denoting the
quadartic Casimir of the each of the fields, the sum being over the
representation of fields that enter $R_i$, and $S$ fully specified
by the respective wavefunction anomalous dimensions.  
It follows that
there are two fixed point values for each coupling:
$R_i^*=0$, or
the non-trivial fixed point solution 
\begin{equation}\label{lepristarequation}
R_i^*=\sum_j (S^{-1})_{ij} (r_j+b_3).
\end{equation}
We shall be interested in the fixed-point solutions of
the couplings $\lambda_{233}, \, \lambda'_{333}, \, h_b, \, h_t$ only,
and shall not consider the $h_\tau$ coupling.  
Therefore, we replace it, as we did earlier, 
by $h_\tau(m_t)=0.6 h_b(m_t)$ at the weak scale in the
determination of the fixed point solutions (\ref{lepristarequation}).
The anomalous dimensions matrix can then be written as:
\begin{equation}
S=\left(
\begin{array}{c c c c}
4 & 3 & 4\eta & 0 \\
1 & 6 & 6+\eta & 1 \\
0 & 6 & 6+\eta & 1 \\
0 & 1 & 1 & 6 \\
\end{array}
\right)
\end{equation}
This leads to the  fixed point values for the ratios:
\begin{eqnarray}
& \displaystyle R_1^* \equiv R^* = 0, & \nonumber \\
& \displaystyle R_2^* \equiv R'^*={315+194 \eta \over 366 \eta -315}, &
\nonumber \\
& \displaystyle R_3^* \equiv R_b^*=-{140\over 122 \eta -105}, & \\
& \displaystyle R_4^* \equiv R_t^* = {110 \eta-105\over 366 \eta-315}. &
\nonumber 
\end{eqnarray}
We note that $R_2^*\equiv R'^*<0$, and therefore, this fixed point
solution is not an acceptable fixed point.  We, thus,  see that a 
simultaneous fixed point for the lepton number violating couplings 
$\lambda_{233}, \, \lambda'_{333}$, and $h_b, \, h_t$
does not exist.

We now consider the two L-violating couplings separately, i.e.,
we shall take either $\lambda_{233}\ll\lambda^{'}_{333}$, or
$\lambda^{'}_{333}\ll\lambda_{233}$, respectively.  In the case
when $\lambda'_{333}$ is the dominant of the couplings, we order
the couplings as $R_i = (R', \, R_b, \, R_t)$, so that the matrix $S$
that enters Eq.~(\ref{lepristarequation}) for this case can be
written as
\begin{equation}
S=\left(
\begin{array}{c c c}
6 & 6+\eta & 1 \\
6 & 6+\eta & 1 \\
1 & 1 & 6 \\
\end{array}
\right).
\end{equation}
Since the determinant of this  matrix  vanishes, there 
are no fixed points in this case.  We thus conclude that a simultaneous
non-zero fixed point for the coupling $\lambda'_{233}, \, h_b, \, h_t$
does not exist. We note that the vanishing of the determinant
corresponds to a solution with a fixed line or  surface.

If $h_b$ is small (e.g., for the case of small $\tan\beta$)
we may reorder the couplings  $R_i = (R_b, \, R', \, R_t)$, and the matrix
$S$, to find the fixed point solution
\begin{eqnarray}
& \displaystyle R_1^*\equiv R_b^*=0, & \nonumber \\
& \displaystyle R_2^*\equiv R'^*={1\over 3}, & \label{2ndlepsolution} \\
& \displaystyle R_3^*\equiv R_t^*={1\over 3}. & \nonumber
\end{eqnarray}
In order to study the stability of this solution, we must obtain
the behaviour of the coupling $R_b^*$ around the origin from
the eigenvalue
\begin{equation}\label{lepsecondevequation}
\lambda_1={1\over b_3}\left[ \sum_{j=2}^3 S_{1j} R_j^* -(r_1+b_3)\right],
\end{equation}
where $r_1=16/3$.  Inserting the relevant $R^*_i$'s 
into (\ref{lepsecondevequation}), we get 
\begin{equation}
\lambda_1=0,
\end{equation}
from which we conclude that the fixed point Eq.~(\ref{2ndlepsolution})
will never be reached 
in the infra-red region. This fixed point is either a saddle point or an
ultra-violet fixed point. We conclude that there are no non-trivial stable
fixed points in the infra-red region for the lepton number violating 
coupling $\lambda^{'}_{333}$.

Finally, we consider the case when $\lambda^{'}_{333} \ll 
\lambda_{233}$. We find the fixed point
solution
\begin{eqnarray}
& \displaystyle R_1^* \equiv R^* = {-315 - 194 \eta\over 12(35 + 6 \eta)}, & 
\nonumber\\
& \displaystyle R_2^* \equiv R_b^* ={  35\over 3(35 + 6 \eta)},  & \\         
& \displaystyle R_3^* \equiv R_t^* = {7(5 +  \eta)\over 3(35 + 6 \eta)},  
& \nonumber      
\end{eqnarray}
which is unphysical. We, therefore, try a fixed point with 
$R_b^* = 0$. We find
\begin{eqnarray}
& \displaystyle R_1^* \equiv R_b^* = 0,    & \nonumber \\
& \displaystyle R_2^* \equiv R^* =  -3/4,  & \\                   
& \displaystyle R_3^* \equiv R_t^* = 7/18,  & \nonumber
\end{eqnarray}
which, again, is unphysical. We have also checked
that: (1) trivial fixed points for 
$\lambda_{233}$ and $h_b$ and
the Pendleton-Ross type fixed point for the top-quark Yukawa coupling, or 
(2) trivial fixed points for $\lambda'_{333}$ and $h_b$ and
the Pendleton-Ross fixed point for the top-quark Yukawa coupling,
are,  both,  unstable in the infra-red region.
We, therefore,  conclude  that there are no fixed point 
solutions for the lepton number violating coupling $\lambda_{233}$.

To summarize, we have  analyzed the one-loop
renormalization group equations for the evolution of Yukawa couplings
in MSSM  with $R_p$ violating couplings 
to the heaviest generation, taking into account 
B and L violating couplings one at a time.
The analysis of the system with $R_p$, and the baryon number,
violating coupling $\lambda''_{233}$ 
yields the surprising and important result that  
only the simultaneous non-trivial
fixed point for this coupling and the top-quark and b-quark
Yukawa couplings $h_t$ and $h_b$ is stable in the infra-red region.
However, the fixed point value for the top-quark coupling here is 
lower than its corresponding value in the MSSM, 
and is incompatible with the measured value of the top-quark mass. 
The $R_p$ conserving solution with $\lambda''_{233}$ attaining its
trivial fixed point, with $h_t$ and $h_b$ attaining non-trivial
fixed points, is infra-red unstable, as is the case for 
trivial fixed points for $\lambda''_{233}$ and $h_b$, with
a non-trivial fixed point for $h_t$.  Our analysis
shows that the usual Pendelton-Ross
type infra-red fixed point of MSSM is unstable in the presence of 
$R_p$ violation, though for small, but negligible,  values of 
$h_b$ and $\lambda^{''}_{233}$ it could be stable.
The system with $L$, and $R_p$,  violating couplings 
does not possess a set of non-trivial
fixed points that are infra-red stable.
Our results
are the first in placing strong theoretical constraints on the nature
of $R_p$ violating couplings from fixed-point and stability
considerations: the fixed points that are unstable, or the 
fixed point that is a saddle point, cannot be realized in 
the infra-red region. The fixed points obtained in this work
are the true fixed points, in contrast to the quasi-fixed points
of~\cite{barger}, and serve as
a lower bound on the relevant $R_p$ violating Yukawa couplings.
In particular, from our analysis of the simultaneous (stable) fixed
point for the baryon number violating coupling $\lambda^{''}_{233}$
and the top and bottom Yukawa couplings, we infer a lower bound
on $\lambda^{''}_{233} \stackrel{>}{\sim} 0.98$.

{\it Note addded}:  After this paper was submitted for publication, 
another paper~\cite{dreiner2} which considers the question of infra-red 
fixed points in the supersymmetric standard model with $R_p$ violation
has appeared.  The fixed points for
$\lambda^{'}_{333}$,  $\lambda^{''}_{233}$ and $h_t$, neglecting all other
Yukawa couplings, is considered. Their results, where there is an overlapp,
agree with ours.  However, unlike in the present work, the stability of the 
fixed points has not been
considered in ~\cite{dreiner2}.
 
\begin{small}
\noindent {\bf Acknowledgements:}
One of the authors (PNP) thanks the Theory Group at DESY
for hospitality while this work was done. 
The work of PNP is supported by the 
University Grants Commission, India under the project 
No. 10-26/98(SR-I).
\end{small}

\end{document}